\documentstyle[12pt,epsfig,graphics,cite]{article}
\begin{document}\bibliographystyle{plain}\begin{titlepage}
\renewcommand{\thefootnote}{\fnsymbol{footnote}}\hfill\begin{tabular}{l}
HEPHY-PUB 805/05\\hep-th/0507281\\July
2005\end{tabular}\\[2cm]\huge\begin{center}{\bf INSTANTANEOUS BETHE--SALPETER
EQUATION WITH EXACT PROPAGATORS}\\\vspace{2cm}
\large{\bf Wolfgang LUCHA\footnote[1]{\normalsize\ {\em E-mail address\/}:
wolfgang.lucha@oeaw.ac.at}}\\[.3cm]\normalsize Institut f\"ur
Hochenergiephysik,\\\"Osterreichische Akademie der
Wissenschaften,\\Nikolsdorfergasse 18, A-1050 Wien,
Austria\\[0.7cm]\large{\bf Franz F.~SCH\"OBERL\footnote[2]{\normalsize\ {\em
E-mail address\/}: franz.schoeberl@univie.ac.at}}\\[.3cm]\normalsize Institut
f\"ur Theoretische Physik, Universit\"at Wien,\\Boltzmanngasse 5, A-1090
Wien, Austria\vfill{\normalsize\bf Abstract}\end{center}\normalsize
Consequent application of the instantaneous approximation to both the
interaction {\em and\/} all propagators of the bound-state constituents
allows us to forge, within the framework~of the Bethe--Salpeter formalism for
the description of bound states, an instantaneous form of the Bethe--Salpeter
equation with exact (i.e., full) propagators of the bound-state constituents.
This instantaneous equation generalizes the well-known Salpeter equation the
derivation of which needs the additional assumption of {\em free\/}
propagation of the bound-state constituents.\vspace{1cm}

\noindent{\em PACS numbers\/}: 11.10.St, 03.65.Ge, 03.65.Pm
\renewcommand{\thefootnote}{\arabic{footnote}}\end{titlepage}

\section{Introduction}Within relativistic quantum field theories, the
appropriate framework, and a standard tool, for the description of bound
states from first principles is the Bethe--Salpeter formalism \cite{BSE}. In
order to circumvent its complexity and problems of interpretation of (all)
its solutions,~a simplification of the Bethe--Salpeter equation has been
sought and soon found: the~Salpeter equation, obtained by a three-dimensional
reduction \cite{SE}. Since then, this equation has~been frequently employed
to study, e.g., the phenomenology of mesons as bound states of~quarks
\cite{Lucha91,Lucha92} either entirely numerically
\cite{LeYaouanc85,Lagae92a,Lagae92b,Resag94,Muenz94,Muenz95,Resag95,Zoeller95,
Klempt95,Muenz96,Parramore95,Parramore96,Olsson95,Olsson96,Koll00,Ricken00,
Merten01,Ricken03} or analytically to the utmost possible extent
\cite{Lucha00:IBSEm=0,Lucha00:IBSE-C4,Lucha00:IBSEnzm,Lucha01:IBSEIAS}.

The Bethe--Salpeter equation reduces to Salpeter's equation under two
assumptions \cite{SE}:\begin{itemize}\item Every bound-state constituent
propagates as a free particle with some effective mass; this constant should
adequately parametrize all dynamical self-energy
contributions.\vspace{-0.95ex}\item All interactions between the particles
which constitute or form the bound state under investigation are
instantaneous in the center-of-momentum frame of the bound state. (Lorentz
covariance is achieved by requiring all interactions to depend, in momentum
space, only on those components of the relative momenta of the involved
bound-state constituents which are perpendicular to the total momentum of
this bound~state \cite{Resag94}.)\end{itemize}However, the above assumption
of free propagators for the bound-state constituents faces a serious
conceptual problem
\cite{Boehm75,Alabiso76,Alabiso77,Pagels77,Hofsaess78,Roth79,Lucha91}, which
provides an argument strongly in favour of the use of the corresponding exact
propagators. In quantum field theory, the Dyson--Schwinger equations connect
these propagators, which are 2-point Green functions, and the particular
$n$-point Green functions that represent the interactions in the
Bethe--Salpeter equation. As consequence of this, the propagators and
interactions cannot be chosen independently from each other. Accordingly, in
non-Abelian gauge theories, such as quantum chromodynamics, the simultaneous
assumptions of free-particle propagation of bound-state constituents and
confining interaction, induced by quantum chromodynamics, are intrinsically
inconsistent. In spite of this, apart from the study of the spectroscopy of
light mesons in Ref.~\cite{Lagae92b}~all~of the above investigations adhere
to the free-propagator approximation from the very beginning.

In view of this, we demonstrate in this note that the free-propagator
assumption can be easily skipped in our construction of a consistent
instantaneous Bethe--Salpeter formalism. We give the resulting relativistic
wave equation for exact albeit instantaneous~propagators.

\section{Instantaneous Bethe--Salpeter Formalism}

\subsection{Homogeneous Bethe--Salpeter Equation}By approximating the
interactions responsible for the formation of the bound states under
consideration by their static limits but retaining---in contrast to the
additional assumption adopted in the derivation of the Salpeter equation
\cite{SE}---the exact form of the propagators of the bound-state
constituents, we construct in this section an instantaneous Bethe--Salpeter
formalism representing a formal improvement of the Salpeter equation. Our
intention~is~to employ the resulting eigenvalue equation of motion eventually
for the description of mesons regarded as bound states of a quark and an
antiquark formed by the strong interactions~\cite{Lucha91}. Accordingly, let
us start the present analysis with a brief exposition of the essential
features of the Bethe--Salpeter equation for bound states composed of a
fermion and an antifermion.

First of all, we have to recall the standard kinematics used for a system of
two particles (labelled $i=1,2$) located at coordinates $x_1,x_2$ and
carrying momenta $p_1,p_2.$ Introducing two arbitrary real (dimensionless)
parameters $\eta_1,\eta_2$ satisfying $\eta_1+\eta_2=1,$ we define for this
two-particle system the center-of-momentum coordinate $X$ and the relative
coordinate $x,$ as well as the conjugate variables, the total momentum $P$
and the relative momentum~$p,$~by$$X\equiv\eta_1\,x_1+\eta_2\,x_2\ ,\quad
x\equiv x_1-x_2\ ,\quad P\equiv p_1+p_2\ ,\quad
p\equiv\eta_2\,p_1-\eta_1\,p_2\ ,$$such that $P\,X+p\,x=p_1\,x_1+p_2\,x_2.$

Within the framework of the Bethe--Salpeter formalism, a bound state of
momentum $P$ and mass $M_{\rm B},$ denoted generically by the Hilbert-space
vector $|{\rm B}(P)\rangle,$ is represented by its Bethe--Salpeter amplitude
$\Psi.$ In the more convenient momentum-space representation, the
Bethe--Salpeter amplitude $\Psi(p)$ is defined as Fourier transform of the
matrix element~of the time-ordered product of the two field operators that
describe the bound-state constituents, evaluated with respect to vacuum state
$|0\rangle$ and bound state $|{\rm B}(P)\rangle,$ after factorizing off the
center-of-momentum motion. Consequently, suppressing all spinor (or Dirac)
and internal indices [as well as the dependence of $\Psi(p)$ on the momentum
$P$ of the bound state $|{\rm B}(P)\rangle$], the momentum-space
Bethe--Salpeter amplitude of any fermion--antifermion bound
state~is$$\Psi(p)\equiv\exp({\rm i}\,P\,X)\int{\rm d}^4x\,\exp({\rm
i}\,p\,x)\,\langle 0|{\rm T}(\psi_1(x_1)\,\bar\psi_2(x_2))|{\rm B}(P)\rangle\
;$$here it is evidently understood that the field operators $\psi_1,$
$\bar\psi_2,$ respectively, are evaluated~at$$x_1=X+\eta_2\,x\ ,\quad
x_2=X-\eta_1\,x\ .$$

The Bethe--Salpeter amplitude $\Psi$ satisfies the formally exact
Bethe--Salpeter equation, which involves two dynamical ingredients: the exact
(or ``full'' or ``dressed'') propagators~of the two bound-state constituents
and the Bethe--Salpeter interaction kernel $K.$ This kernel is a fully
amputated 4-point Green function, defined (merely perturbatively!) as the sum
of the countable infinity of all Feynman diagrams for two-particle into
two-particle scattering. Moreover, it has to be Bethe--Salpeter irreducible,
i.e., two-particle irreducible with respect to the two bound-state
constituents. In the momentum space the kernel $K(p,q)$ depends~on the
relative momenta $p$ and $q$ of initial and final state (as well as on the
momentum $P$ of the bound state). Denoting the propagator of the fermion
$i=1,2$ by $S_i(p),$ the Bethe--Salpeter equation for a fermion--antifermion
bound state reads, in momentum-space representation,
\begin{equation}S_1^{-1}(p_1)\,\Psi(p)\,S_2^{-1}(-p_2)=\frac{{\rm
i}}{(2\pi)^4}\int{\rm d}^4q\,K(p,q)\,\Psi(q)\ .\label{Eq:BSE}\end{equation}
Multiplying Eq.~(\ref{Eq:BSE}) by $S_1(p_1)$ from the left and $S_2(-p_2)$
from the right, the Bethe--Salpeter equation becomes
\begin{equation}\Psi(p)=\frac{{\rm i}}{(2\pi)^4}\,S_1(p_1)\int{\rm
d}^4q\,K(p,q)\,\Psi(q)\,S_2(-p_2)\ .\label{Eq:BSE1}\end{equation}

\subsection{Instantaneous Approximation}The instantaneous (or static)
approximation to the Bethe--Salpeter formalism is based on the assumption
that the Bethe--Salpeter kernel $K$ entering in the Bethe--Salpeter equations
(\ref{Eq:BSE}) or (\ref{Eq:BSE1}) depends only on the spatial components
$\mbox{\boldmath{$p$}}$ and $\mbox{\boldmath{$q$}}$ of the relative momenta
$p$~and~$q$:\begin{equation}K(p,q)=\hat
K(\mbox{\boldmath{$p$}},\mbox{\boldmath{$q$}})\
.\label{Eq:IA}\end{equation}This restriction corresponds to the assumption of
an instantaneous interaction between the bound-state constituents and amounts
to ignoring all retardation effects in the interaction.

In the instantaneous approximation the Bethe--Salpeter equation may be
reduced to an equation of motion for a Salpeter amplitude (or equal-time
bound-state wave function) $\Phi.$ In momentum-space representation the
Salpeter amplitude $\Phi(\mbox{\boldmath{$p$}})$ is obtained by integrating
the Bethe--Salpeter amplitude $\Psi(p)$ over the time component $p_0$ of the
relative momentum:$$\Phi(\mbox{\boldmath{$p$}})\equiv\frac{1}{2\pi}\int{\rm
d}p_0\,\Psi(p)\ .$$Introducing, for the sake of notational simplicity, for
the interaction term the abbreviation
$$I(\mbox{\boldmath{$p$}})\equiv\frac{1}{(2\pi)^3}\int{\rm d}^3q\,\hat
K(\mbox{\boldmath{$p$}},\mbox{\boldmath{$q$}})\,\Phi(\mbox{\boldmath{$q$}})\
,$$upon integration over $p_0$ the Bethe--Salpeter equation (\ref{Eq:BSE1})
assumes the instantaneous form
\begin{equation}\Phi(\mbox{\boldmath{$p$}})=\frac{{\rm i}}{2\pi}\int{\rm
d}p_0\,S_1(p_1)\,I(\mbox{\boldmath{$p$}})\,S_2(-p_2)\
;\label{Eq:BSE-I}\end{equation}here the momenta $p_1,p_2$ are, of course,
regarded as functions of total and relative momenta:$$p_1=\eta_1\,P+p\ ,\quad
p_2=\eta_2\,P-p\ .$$

\subsection{Exact Fermion Propagator}\label{Subsec:EFP}The actual evaluation
of the integral over the time component $p_0$ of the relative momentum $p$ of
the tensor product of fermion propagators in the (integrated) Bethe--Salpeter
equation (\ref{Eq:BSE-I}) requires the knowledge of the dependence of the
(full) fermion propagator $S_i(p)$ on $p_0.$

The fermion propagator $S_i(p)$ is the solution of the fermion
Dyson--Schwinger equation, in view of its r\^ole in all analyses of the
phenomenon of dynamical chiral-symmetry breaking also called the gap equation
(for reviews of this subject consult, for instance, Refs.~\cite{Roberts94,
Roberts00a,Roberts00b,Alkofer01,Maris02,Maris03,Roberts03}). By Lorentz
covariance (if preserved by gauge fixing) the general solution $S_i(p)$ for
the~exact fermion propagator is represented by only two Lorentz-scalar
functions, $M_i(p^2)$ and $Z_i(p^2)$:\begin{equation}S_i(p)=\frac{{\rm
i}\,Z_i(p^2)}{\not\!p-M_i(p^2)+{\rm i}\,\varepsilon}\ ,\quad\not\!p\equiv
p^\mu\,\gamma_\mu\ .\label{Eq:exactFP}\end{equation}Hence, the exact
propagator $S_i(p)$ is related to the free propagator of a fermion of
mass~$m,$\begin{equation}S_0(p,m)=\frac{{\rm i}}{\not\!p-m+{\rm
i}\,\varepsilon}\equiv{\rm i}\,\frac{\not\!p+m}{p^2-m^2+{\rm
i}\,\varepsilon}\ ,\label{Eq:freeFP}\end{equation}according to
$S_i(p)=Z_i(p^2)\,S_0(p,M_i(p^2)).$ This form of the exact fermion
propagators $S_i(p)$ is equivalent to an alternate representation in terms of
scalar functions $A_i(p^2)$ and~$B_i(p^2),$$$S_i(p)=\frac{{\rm
i}}{A_i(p^2)\!\not\!p-B_i(p^2)+{\rm i}\,\varepsilon}\ ,$$by means of the
identifications$$M_i(p^2)=\frac{B_i(p^2)}{A_i(p^2)}\ ,\quad
Z_i(p^2)=\frac{1}{A_i(p^2)}\ .$$

Now, in order to continue our present quest for a generalization of the
Salpeter equation we have to impose a restriction on the two Lorentz-scalar
functions, denoted by $M_i(p^2)$ and $Z_i(p^2)$ in the representation
(\ref{Eq:exactFP}), which characterize each exact fermion propagator
$S_i(p).$ In full accordance with the spirit of the instantaneous
approximation to the interaction kernel entering in the Bethe--Salpeter
equation, as well as for the sake of conceptual simplicity, we neglect
retardation effects also in the propagators and assume both of the functions
$M_i(p^2)$ and $Z_i(p^2)$ to depend only on the spatial components,
$\mbox{\boldmath{$p$}},$ of the involved momentum $p.$ This means we make the
substitutions $M_i(p^2)\to M_i(\mbox{\boldmath{$p$}}^2),$ $Z_i(p^2)\to
Z_i(\mbox{\boldmath{$p$}}^2),$ which entails for the above-mentioned relation
between exact and free propagator
$S_i(p)=Z_i(\mbox{\boldmath{$p$}}^2)\,S_0(p,M_i(\mbox{\boldmath{$p$}}^2)).$
Our (pragmatic) point of view is supported, for example, within quantum
chromodynamics by findings for the quark propagator
\cite{Finger82,LeYaouanc84,Adler84,Kocic86,Alkofer88,Klevansky88,Alkofer89}
obtained from effective Coulomb-gauge model Hamiltonians designed to
implement effects of spontaneous breakdown of chiral symmetry. Clearly, our
assumption corresponds to the simplest conceivable dependence of $S_i(p)$
on~$p_0.$

\subsection{Exact-Propagator Instantaneous Bethe--Salpeter Formalism}
\label{EP-IBSE}Upon acceptance of the ``$p_0^2=0$'' approximation for both of
the involved propagators~in the sense as specified above in
Subsect.~\ref{Subsec:EFP}, the formulation of the corresponding instantaneous
Bethe--Salpeter equation follows a ``well-paved road.'' We introduce the
one-particle energy$$E_i(\mbox{\boldmath{$p$}})\equiv
\sqrt{\mbox{\boldmath{$p$}}^2+M_i^2(\mbox{\boldmath{$p$}}^2)}\ ,\quad i=1,2\
,$$and the generalized Dirac Hamiltonian$$H_i(\mbox{\boldmath{$p$}})\equiv
\gamma_0\,[\mbox{\boldmath{$\gamma$}}\cdot\mbox{\boldmath{$p$}}
+M_i(\mbox{\boldmath{$p$}}^2)]\ ,\quad i=1,2\ ,$$satisfying
$H_i^2(\mbox{\boldmath{$p$}})=E_i^2(\mbox{\boldmath{$p$}})$ and
$H_i(\mbox{\boldmath{$p$}})\,\gamma_0=\gamma_0\,H_i(-\mbox{\boldmath{$p$}}).$
With these notions, we define~energy projection operators
$\Lambda_i^\pm(\mbox{\boldmath{$p$}})$ for positive or negative energy
$\pm\,E_i(\mbox{\boldmath{$p$}})$ of particle $i,$ $i=1,2,$~by
$$\Lambda_i^\pm(\mbox{\boldmath{$p$}})\equiv\frac{E_i(\mbox{\boldmath{$p$}})\pm
H_i(\mbox{\boldmath{$p$}})}{2\,E_i(\mbox{\boldmath{$p$}})}\ ,\quad i=1,2\
,$$exhibiting all projection-operator properties
$\Lambda_i^\pm(\mbox{\boldmath{$p$}})\,\Lambda_i^\pm(\mbox{\boldmath{$p$}})
=\Lambda_i^\pm(\mbox{\boldmath{$p$}}),$
$\Lambda_i^\pm(\mbox{\boldmath{$p$}})\,\Lambda_i^\mp(\mbox{\boldmath{$p$}})
=0$~and
$\Lambda_i^+(\mbox{\boldmath{$p$}})+\Lambda_i^-(\mbox{\boldmath{$p$}})=1,$
and satisfying $\Lambda_i^\pm(\mbox{\boldmath{$p$}})\,\gamma_0
=\gamma_0\,\Lambda_i^\pm(-\mbox{\boldmath{$p$}}).$ All of this enters in the
identity$$\not\!p+M_i(\mbox{\boldmath{$p$}}^2)=
\left([p_0+E_i(\mbox{\boldmath{$p$}})]\,\Lambda_i^+(\mbox{\boldmath{$p$}})+
[p_0-E_i(\mbox{\boldmath{$p$}})]\,\Lambda_i^-(\mbox{\boldmath{$p$}})\right)
\gamma_0\ .$$With the help of this identity, the exact fermion propagator
$S_i(p)$ may be cast into the~form\begin{equation}S_i(p)={\rm
i}\,Z_i(\mbox{\boldmath{$p$}}^2)\left(
\frac{\Lambda_i^+(\mbox{\boldmath{$p$}})}{p_0-E_i(\mbox{\boldmath{$p$}})+{\rm
i}\,\varepsilon}+
\frac{\Lambda_i^-(\mbox{\boldmath{$p$}})}{p_0+E_i(\mbox{\boldmath{$p$}})-{\rm
i}\,\varepsilon}\right)\gamma_0\ .\label{Eq:FP}\end{equation}This is nothing
else but the partial fraction decomposition of the right-hand side of
Eq.~(\ref{Eq:exactFP}).

With the decomposition (\ref{Eq:FP}) of the fermion propagators at one's
disposal, the evaluation of the integral of the tensor product of fermion
propagators in the Bethe--Salpeter equation (\ref{Eq:BSE-I}) is
straightforward. Employing the residue theorem, a simple contour integration
yields\begin{eqnarray*}&&\int{\rm d}p_0\,S_1(p_1)\otimes S_2(-p_2)\\[1ex]&&=-
\,2\pi\,{\rm i}\,Z_1(\mbox{\boldmath{$p$}}_1^2)\,Z_2(\mbox{\boldmath{$p$}}_2^2)
\left(\frac{\Lambda_1^+(\mbox{\boldmath{$p$}}_1)\,\gamma_0\otimes\gamma_0\,
\Lambda_2^-(\mbox{\boldmath{$p$}}_2)}
{P_0-E_1(\mbox{\boldmath{$p$}}_1)-E_2(\mbox{\boldmath{$p$}}_2)}
-\frac{\Lambda_1^-(\mbox{\boldmath{$p$}}_1)\,\gamma_0\otimes\gamma_0\,
\Lambda_2^+(\mbox{\boldmath{$p$}}_2)}
{P_0+E_1(\mbox{\boldmath{$p$}}_1)+E_2(\mbox{\boldmath{$p$}}_2)}\right).
\end{eqnarray*}By use of this result in Eq.~(\ref{Eq:BSE-I}), we arrive at
our instantaneous Bethe--Salpeter equation for fermion--antifermion bound
states, with exact propagators of the bound-state
constituents:\begin{equation}\Phi(\mbox{\boldmath{$p$}})
=Z_1(\mbox{\boldmath{$p$}}_1^2)\,Z_2(\mbox{\boldmath{$p$}}_2^2)
\left(\frac{\Lambda_1^+(\mbox{\boldmath{$p$}}_1)\,\gamma_0\,
I(\mbox{\boldmath{$p$}})\,\gamma_0\,\Lambda_2^-(\mbox{\boldmath{$p$}}_2)}
{P_0-E_1(\mbox{\boldmath{$p$}}_1)-E_2(\mbox{\boldmath{$p$}}_2)}
-\frac{\Lambda_1^-(\mbox{\boldmath{$p$}}_1)\,\gamma_0\,
I(\mbox{\boldmath{$p$}})\,\gamma_0\,\Lambda_2^+(\mbox{\boldmath{$p$}}_2)}
{P_0+E_1(\mbox{\boldmath{$p$}}_1)+E_2(\mbox{\boldmath{$p$}}_2)}\right).
\label{Eq:IBSE}\end{equation}Upon applying to Eq.~(\ref{Eq:IBSE})
$\Lambda_1^\pm(\mbox{\boldmath{$p$}}_1)$ from the left and
$\Lambda_2^\pm(\mbox{\boldmath{$p$}}_2)$ from the right, one realizes that
precisely half of the components of the Salpeter amplitude
$\Phi(\mbox{\boldmath{$p$}})$ have to vanish
identically:\begin{equation}\Lambda_1^+(\mbox{\boldmath{$p$}}_1)\,
\Phi(\mbox{\boldmath{$p$}})\,\Lambda_2^+(\mbox{\boldmath{$p$}}_2)
=\Lambda_1^-(\mbox{\boldmath{$p$}}_1)\,\Phi(\mbox{\boldmath{$p$}})\,
\Lambda_2^-(\mbox{\boldmath{$p$}}_2)=0\ .\label{Eq:constraint}\end{equation}
The difference of these two constraints on the Salpeter amplitude
$\Phi(\mbox{\boldmath{$p$}})$
yields$$\frac{H_1(\mbox{\boldmath{$p$}}_1)}{E_1(\mbox{\boldmath{$p$}}_1)}
\,\Phi(\mbox{\boldmath{$p$}})+\Phi(\mbox{\boldmath{$p$}})\,
\frac{H_2(\mbox{\boldmath{$p$}}_2)}{E_2(\mbox{\boldmath{$p$}}_2)}=0\ .$$

\subsection{Formulation as Equivalent Eigenvalue Problem}Starting already
with the pioneering work of Salpeter \cite{SE}, it has become a common
practice to cast the instantaneous Bethe--Salpeter equation into the shape of
an eigenvalue problem. Let us perform this step in the present study too:
with the help of, for instance, the identity
$$\Lambda_1^\pm(\mbox{\boldmath{$p$}}_1)
\,[H_1(\mbox{\boldmath{$p$}}_1)\,\Phi(\mbox{\boldmath{$p$}})
-\Phi(\mbox{\boldmath{$p$}})\,H_2(\mbox{\boldmath{$p$}}_2)]\,
\Lambda_2^\mp(\mbox{\boldmath{$p$}}_2)
=\pm\,[E_1(\mbox{\boldmath{$p$}}_1)+E_2(\mbox{\boldmath{$p$}}_2)]\,
\Lambda_1^\pm(\mbox{\boldmath{$p$}}_1)\,\Phi(\mbox{\boldmath{$p$}})\,
\Lambda_2^\mp(\mbox{\boldmath{$p$}}_2)\ ,$$which follows from
$$\Lambda_i^\pm(\mbox{\boldmath{$p$}})\,H_i(\mbox{\boldmath{$p$}})
=H_i(\mbox{\boldmath{$p$}})\,\Lambda_i^\pm(\mbox{\boldmath{$p$}})
=\pm\,E_i(\mbox{\boldmath{$p$}})\,\Lambda_i^\pm(\mbox{\boldmath{$p$}})\
,\quad i=1,2\ ,$$we easily extract from Eq.~(\ref{Eq:IBSE}) our (explicit)
eigenvalue equation for the Salpeter amplitude $\Phi(\mbox{\boldmath{$p$}}),$
with the (total) energy $P_0$ of the bound state $|{\rm B}(P)\rangle$ as the
corresponding eigenvalue:
\begin{eqnarray}&&H_1(\mbox{\boldmath{$p$}}_1)\,\Phi(\mbox{\boldmath{$p$}})
-\Phi(\mbox{\boldmath{$p$}})\,H_2(\mbox{\boldmath{$p$}}_2)\nonumber\\[1ex]
&&+\,Z_1(\mbox{\boldmath{$p$}}_1^2)\,Z_2(\mbox{\boldmath{$p$}}_2^2)\,
[\Lambda_1^+(\mbox{\boldmath{$p$}}_1)\,\gamma_0\,I(\mbox{\boldmath{$p$}})\,
\gamma_0\,\Lambda_2^-(\mbox{\boldmath{$p$}}_2)
-\Lambda_1^-(\mbox{\boldmath{$p$}}_1)\,\gamma_0\,I(\mbox{\boldmath{$p$}})\,
\gamma_0\,\Lambda_2^+(\mbox{\boldmath{$p$}}_2)]\nonumber\\[1ex]
&&=P_0\,\Phi(\mbox{\boldmath{$p$}})\ .\label{Eq:EVE}\end{eqnarray}It goes
without saying that every solution of the above eigenvalue equation for the
Salpeter amplitude $\Phi(\mbox{\boldmath{$p$}})$ has to be still subject to
the constraints (\ref{Eq:constraint}). More precisely, this eigenvalue
equation (\ref{Eq:EVE}) alone is definitely {\em not\/} equivalent to the
``true'' instantaneous Bethe--Salpeter equation (\ref{Eq:IBSE}). This fact
can be revealed by application of the projectors
$\Lambda_1^\pm(\mbox{\boldmath{$p$}}_1)$ from~the left and the projectors
$\Lambda_2^\pm(\mbox{\boldmath{$p$}}_2)$ from the right to the instantaneous
Bethe--Salpeter equation~(\ref{Eq:IBSE}) and the eigenvalue equation
(\ref{Eq:EVE}): The $\Lambda_1^\pm(\mbox{\boldmath{$p$}}_1)\otimes
\Lambda_2^\mp(\mbox{\boldmath{$p$}}_2)$ components of Eq.~(\ref{Eq:IBSE}) and
Eq.~(\ref{Eq:EVE}) are certainly identical whereas the
$\Lambda_1^\pm(\mbox{\boldmath{$p$}}_1)\otimes
\Lambda_2^\pm(\mbox{\boldmath{$p$}}_2)$ projections yield, for the
instantaneous Bethe--Salpeter equation (\ref{Eq:IBSE}), the two constraints
(\ref{Eq:constraint}) but, for the eigenvalue equation (\ref{Eq:EVE}),
$$\{P_0\mp[E_1(\mbox{\boldmath{$p$}}_1)-E_2(\mbox{\boldmath{$p$}}_2)]\}\,
\Lambda_1^\pm(\mbox{\boldmath{$p$}}_1)\,\Phi(\mbox{\boldmath{$p$}})\,
\Lambda_2^\pm(\mbox{\boldmath{$p$}}_2)=0\ ,$$which does not necessarily imply
$\Lambda_1^\pm(\mbox{\boldmath{$p$}}_1)\,\Phi(\mbox{\boldmath{$p$}})\,
\Lambda_2^\pm(\mbox{\boldmath{$p$}}_2)=0.$ Consequently, equivalence~only
holds between the instantaneous Bethe--Salpeter equation (\ref{Eq:IBSE}), on
the one hand, and the set formed by the eigenvalue equation (\ref{Eq:EVE})
{\em amended\/} by the constraints (\ref{Eq:constraint}), on the other~hand.

In the rest frame of the two-particle system under consideration, defined by
$\mbox{\boldmath{$P$}}={\bf 0},$ the time component of the total momentum $P$
reduces to the bound-state mass eigenvalue~$M_{\rm B},$ that is, $P_0=M_{\rm
B}.$ The instantaneous Bethe--Salpeter equation, in particular, its form
given by the eigenvalue equation (\ref{Eq:EVE}), then involves only the
relative momentum
$\mbox{\boldmath{$p$}}=\mbox{\boldmath{$p$}}_1=-\mbox{\boldmath{$p$}}_2$:
\begin{eqnarray}&&H_1(\mbox{\boldmath{$p$}})\,\Phi(\mbox{\boldmath{$p$}})
-\Phi(\mbox{\boldmath{$p$}})\,H_2(-\mbox{\boldmath{$p$}})\nonumber\\[1ex]
&&+\,Z_1(\mbox{\boldmath{$p$}}^2)\,Z_2(\mbox{\boldmath{$p$}}^2)\,
[\Lambda_1^+(\mbox{\boldmath{$p$}})\,\gamma_0\,I(\mbox{\boldmath{$p$}})\,
\Lambda_2^-(\mbox{\boldmath{$p$}})\,\gamma_0-
\Lambda_1^-(\mbox{\boldmath{$p$}})\,\gamma_0\,I(\mbox{\boldmath{$p$}})\,
\Lambda_2^+(\mbox{\boldmath{$p$}})\,\gamma_0]\nonumber\\[1ex]&&=M_{\rm
B}\,\Phi(\mbox{\boldmath{$p$}})\ .\label{Eq:EVE-CMS}\end{eqnarray}

\subsection{Slightly Alternate Definition of Bethe--Salpeter Amplitude}In
order to make contact with some notation adopted in investigations of the
instantaneous Bethe--Salpeter equation presented in
Refs.~\cite{Lagae92a,Lagae92b,Olsson95,Olsson96} which construct the
Bethe--Salpeter amplitude from the matrix element $\langle 0|{\rm
T}(\psi_1(x_1)\,\psi_2^\dagger(x_2))|{\rm B}(P)\rangle,$ i.e., using
$\psi_2^\dagger(x)$ instead of $\bar\psi_2(x),$ we have to redefine both
Salpeter amplitude and Bethe--Salpeter kernel according~to
\begin{equation}\chi(\mbox{\boldmath{$p$}})\equiv\Phi(\mbox{\boldmath{$p$}})\,
\gamma_0\ ,\quad
W(\mbox{\boldmath{$p$}},\mbox{\boldmath{$q$}})\,\chi(\mbox{\boldmath{$q$}})
\equiv\gamma_0\,\hat
K(\mbox{\boldmath{$p$}},\mbox{\boldmath{$q$}})\,\Phi(\mbox{\boldmath{$q$}})\
,\label{Eq:AK-redef}\end{equation}and, for convenience, we introduce for the
instantaneous interaction term the
abbreviation$$J(\mbox{\boldmath{$p$}})\equiv\gamma_0\,I(\mbox{\boldmath{$p$}})
=\frac{1}{(2\pi)^3}\int{\rm d}^3q\,
W(\mbox{\boldmath{$p$}},\mbox{\boldmath{$q$}})\,\chi(\mbox{\boldmath{$q$}})\
.$$With this, we obtain the generalization, to the case of exact propagators
of the bound-state constituents, of the eigenvalue form of the Salpeter
equation recalled in Refs.~\cite{Lagae92a,Resag94,Muenz94,Muenz96,Olsson95}:
\begin{eqnarray*}&&H_1(\mbox{\boldmath{$p$}})\,\chi(\mbox{\boldmath{$p$}})
-\chi(\mbox{\boldmath{$p$}})\,H_2(\mbox{\boldmath{$p$}})
+Z_1(\mbox{\boldmath{$p$}}^2)\,Z_2(\mbox{\boldmath{$p$}}^2)\,
[\Lambda_1^+(\mbox{\boldmath{$p$}})\,J(\mbox{\boldmath{$p$}})\,
\Lambda_2^-(\mbox{\boldmath{$p$}})
-\Lambda_1^-(\mbox{\boldmath{$p$}})\,J(\mbox{\boldmath{$p$}})\,
\Lambda_2^+(\mbox{\boldmath{$p$}})]\\[1ex] &&=M_{\rm
B}\,\chi(\mbox{\boldmath{$p$}})\ .\end{eqnarray*}

As a trivial observation, the redefinition (\ref{Eq:AK-redef}) of Salpeter
amplitude and Bethe--Salpeter kernel must be taken into account in the
discussion of the Lorentz (or spin) structure of~the interaction kernel.
Suppose that the kernel $\hat K(\mbox{\boldmath{$p$}},\mbox{\boldmath{$q$}})$
acts on the Salpeter amplitude $\Phi(\mbox{\boldmath{$p$}})$~as$$\hat
K(\mbox{\boldmath{$p$}},\mbox{\boldmath{$q$}})\,\Phi(\mbox{\boldmath{$q$}})=
V(\mbox{\boldmath{$p$}},\mbox{\boldmath{$q$}})\,\Gamma\,
\Phi(\mbox{\boldmath{$q$}})\,\Gamma\ ,$$with some Lorentz-scalar interaction
function $V(\mbox{\boldmath{$p$}},\mbox{\boldmath{$q$}})$ and generic Dirac
matrices $\Gamma.$ In this case, the kernel $\hat
K(\mbox{\boldmath{$p$}},\mbox{\boldmath{$q$}})$ is said to be of Lorentz
structure $\Gamma\otimes\Gamma$ and written symbolically~in the form of the
product of the interaction function
$V(\mbox{\boldmath{$p$}},\mbox{\boldmath{$q$}})$ and the tensor product
$\Gamma\otimes\Gamma$:$$\hat K(\mbox{\boldmath{$p$}},\mbox{\boldmath{$q$}})
=V(\mbox{\boldmath{$p$}},\mbox{\boldmath{$q$}})\,\Gamma\otimes\Gamma\ .$$The
redefined interaction kernel $W(\mbox{\boldmath{$p$}},\mbox{\boldmath{$q$}})$
acts on the redefined Salpeter amplitude $\chi(\mbox{\boldmath{$p$}})$~as
$$W(\mbox{\boldmath{$p$}},\mbox{\boldmath{$q$}})\,\chi(\mbox{\boldmath{$q$}})
=V(\mbox{\boldmath{$p$}},\mbox{\boldmath{$q$}})\,\tilde\Gamma\,
\chi(\mbox{\boldmath{$q$}})\,\tilde\Gamma\ ,$$where the Lorentz structure
$\tilde\Gamma\otimes\tilde\Gamma$ of the redefined kernel
$W(\mbox{\boldmath{$p$}},\mbox{\boldmath{$q$}})
=V(\mbox{\boldmath{$p$}},\mbox{\boldmath{$q$}})\,
\tilde\Gamma\otimes\tilde\Gamma$ is related to its counterpart in the
original kernel $\hat K(\mbox{\boldmath{$p$}},\mbox{\boldmath{$q$}}),$ i.e.,
$\Gamma\otimes\Gamma,$ according to
$\tilde\Gamma\otimes\tilde\Gamma=\gamma_0\,\Gamma\otimes\gamma_0\,\Gamma.$

\subsection{Salpeter Equation}In addition to the instantaneous approximation
(\ref{Eq:IA}), the derivation of the Salpeter equation \cite{SE} invokes the
further assumption that the exact fermion propagators $S_i(p)$ entering in
the Bethe--Salpeter equations (\ref{Eq:BSE}) or (\ref{Eq:BSE1}) may be
reasonably approximated by the corresponding free propagators
(\ref{Eq:freeFP}), that is, $S_i(p)=S_0(p,m_i).$ In this approximation the
mass parameters $m_i$ in the free propagators are then interpreted as some
effective (or ``constituent'') masses of the fermionic bound-state
constituents. Accordingly the Salpeter equation may be easily recovered from
the more general instantaneous Bethe--Salpeter formalism developed in this
analysis by performing, for the exact functions $M_i(p^2)$ and $Z_i(p^2),$
the free-propagator limit$$M_i(p^2)\to m_i\ ,\quad Z_i(p^2)\to 1\ .$$With our
definitions of energies $E_i(\mbox{\boldmath{$p$}})$ and Hamiltonians
$H_i(\mbox{\boldmath{$p$}})$ in Subsect.~\ref{EP-IBSE}, this
implies$$E_i(\mbox{\boldmath{$p$}})\to\sqrt{\mbox{\boldmath{$p$}}^2+m_i^2}\
,\quad H_i(\mbox{\boldmath{$p$}})\to
\gamma_0\,(\mbox{\boldmath{$\gamma$}}\cdot\mbox{\boldmath{$p$}}+m_i)\
.$$Other choices for the effective propagators of the free (i.e.,
non-interacting) fermions in the instantaneous Bethe--Salpeter equation
(\ref{Eq:BSE-I}) entail different three-dimensional reductions of the
Bethe--Salpeter equation. For comparisons of the various proposals, see
Refs.~\cite{Babutsidze98,Babutsidze99,Kopaleishvili01,Babutsidze03}.

\section{Summary, Conclusions, and Outlook}We analyzed the instantaneous
limit of the Bethe--Salpeter formalism and, by applying~this approximation
not only to the interaction kernel but also to the propagators of the
involved bound-state constituents, we formulated, for the case of the exact
(or full) propagators,~the instantaneous Bethe--Salpeter equation for
fermion--antifermion bound states (\ref{Eq:IBSE}) as well as its
representation as eigenvalue problem in both arbitrary reference frame
(\ref{Eq:EVE}) and the rest frame of the bound state (\ref{Eq:EVE-CMS}).
Retaining (an approximation to) the full fermion propagator should facilitate
the embedding of a mechanism for spontaneous chiral symmetry breaking. All
exact propagators required may be extracted, with due care, from lattice
computations. Our findings thus may improve the understanding of various
bound systems; corresponding studies are in progress \cite{Lucha05}. Clearly,
the generalization of our results to systems involving~as bound-state
constituents also particles other than spin-$\frac{1}{2}$ fermions runs along
similar lines.

\section*{Acknowledgements}One of us (W.~L.) would like to thank Craig
D.~Roberts for a lot of stimulating discussions.


\small\begin{thebibliography}{30}
\bibitem{BSE}E.~E.~Salpeter and H.~A.~Bethe, Phys.~Rev.~{\bf 84} (1951) 1232.
\bibitem{SE}E.~E.~Salpeter, Phys.~Rev.~{\bf 87} (1952) 328.
\bibitem{Lucha91}W.~Lucha, F.~F.~Sch\"oberl, and D.~Gromes, Phys.~Rep.~{\bf
200} (1991) 127.
\bibitem{Lucha92}W.~Lucha and F.~F.~Sch\"oberl, Int.~J.~Mod.~Phys.~A {\bf 7}
(1992) 6431.

\newpage

\bibitem{LeYaouanc85}A.~Le Yaouanc, L.~Oliver, S.~Ono, O.~P\`ene, and
J.-C.~Raynal, Phys.~Rev.~D {\bf 31} (1985) 137.
\bibitem{Lagae92a}J.-F.~Laga\"e, Phys.~Rev.~D {\bf 45} (1992) 305.
\bibitem{Lagae92b}J.-F.~Laga\"e, Phys.~Rev.~D {\bf 45} (1992) 317.
\bibitem{Resag94}J.~Resag, C.~R.~M\"unz, B.~C.~Metsch, and H.~R.~Petry,
Nucl.~Phys.~A {\bf 578} (1994) 397, nucl-th/9307026;\\J.~Resag, {\em Analysis
of the instantaneous Bethe--Salpeter equation and its application to $q\bar
q$ bound states\/}, Ph.~D thesis, University of Bonn (1994).
\bibitem{Muenz94}C.~R.~M\"unz, J.~Resag, B.~C.~Metsch, and H.~R.~Petry,
Nucl.~Phys.~A {\bf 578} (1994) 418, nucl-th/9307027;\\C.~R.~M\"unz, {\em
Meson decays and form factors in a relativistic quark model\/}, Ph.~D thesis,
University of Bonn (1994).
\bibitem{Muenz95}C.~R.~M\"unz, J.~Resag, B.~C.~Metsch, and H.~R.~Petry,
Phys.~Rev.~C {\bf 52} (1995) 2110, nucl-th/9406035.
\bibitem{Resag95}J.~Resag and C.~R.~M\"unz, Nucl.~Phys.~A {\bf 590} (1995)
735, nucl-th/9407033.
\bibitem{Zoeller95}G.~Z\"oller, S.~Hainzl, C.~R.~M\"unz, and M.~Beyer,
Z.~Phys.~C {\bf 68} (1995) 103, hep-ph/9412355.
\bibitem{Klempt95}E.~Klempt, B.~C.~Metsch, C.~R.~M\"unz, and H.~R.~Petry,
Phys.~Lett.~B {\bf 361} (1995) 160, hep-ph/9507449.
\bibitem{Muenz96}C.~R.~M\"unz, Nucl.~Phys.~A {\bf 609} (1996) 364,
hep-ph/9601206.
\bibitem{Parramore95}J.~Parramore and J.~Piekarewicz, Nucl.~Phys.~A {\bf 585}
(1995) 705, nucl-th/9402019.
\bibitem{Parramore96}J.~Parramore, H.-C.~Jean, and J.~Piekarewicz,
Phys.~Rev.~C {\bf 53} (1996) 2449, nucl-th/9510024.
\bibitem{Olsson95}M.~G.~Olsson, S.~Veseli, and K.~Williams, Phys.~Rev.~D {\bf
52} (1995) 5141, hep-ph/9503477.
\bibitem{Olsson96}M.~G.~Olsson, S.~Veseli, and K.~Williams, Phys.~Rev.~D {\bf
53} (1996) 504, hep-ph/9504221.
\bibitem{Koll00}M.~Koll, R.~Ricken, D.~Merten, B.~C.~Metsch, and H.~R.~Petry,
Eur.~Phys.~J.~A {\bf 9} (2000)~73, hep-ph/0008220;\\M.~Koll, {\em Electroweak
processes with light mesons in a relativistic quark model\/}, Ph.~D thesis,
University of Bonn (2001).
\bibitem{Ricken00}R.~Ricken, M.~Koll, D.~Merten, B.~Ch.~Metsch, and
H.~R.~Petry, Eur.~Phys.~J.~A {\bf 9} (2000)~221, hep-ph/0008221;\\R.~Ricken,
{\em Properties of light mesons in a relativistic quark model\/}, Ph.~D
thesis, University of Bonn (2001).
\bibitem{Merten01}D.~Merten, R.~Ricken, M.~Koll, B.~Metsch, and H.~Petry,
Eur.~Phys.~J.~A {\bf 13} (2002) 477, hep-ph/0104029;\\D.~Merten, {\em Hadron
form factors and decays\/}, Ph.~D thesis, University of Bonn (2002).
\bibitem{Ricken03}R.~Ricken, M.~Koll, D.~Merten, and B.~Ch.~Metsch,
Eur.~Phys.~J.~A {\bf 18} (2003) 667, hep-ph/0302124.
\bibitem{Lucha00:IBSEm=0}W.~Lucha, K.~Maung Maung, and F.~F.~Sch\"oberl,
Phys.~Rev.~D {\bf 63} (2001) 056002, hep-ph/0009185.
\bibitem{Lucha00:IBSE-C4}W.~Lucha, K.~Maung Maung, and F.~F.~Sch\"oberl, in:
Proc.~Int.~Conf.~on {\em Quark Confinement and the Hadron Spectrum IV\/},
edited by W.~Lucha and K.~Maung Maung (World Scientific, New
Jersey/London/Singapore/Hong Kong, 2002), p.~340, hep-ph/0010078.
\bibitem{Lucha00:IBSEnzm}W.~Lucha, K.~Maung Maung, and F.~F.~Sch\"oberl,
Phys.~Rev.~D {\bf 64} (2001) 036007, hep-ph/0011235.
\bibitem{Lucha01:IBSEIAS}W.~Lucha and F.~F.~Sch\"oberl, Int.~J.~Mod.~Phys.~A
{\bf 17} (2002) 2233, hep-ph/0109165.
\bibitem{Boehm75}M.~B\"ohm, Nucl.~Phys.~B {\bf 91} (1975) 494.
\bibitem{Alabiso76}C.~Alabiso and G.~Schierholz, Nucl.~Phys.~B {\bf 110}
(1976) 81.
\bibitem{Alabiso77}C.~Alabiso and G.~Schierholz, Nucl.~Phys.~B {\bf 126}
(1977) 461.
\bibitem{Pagels77}H.~Pagels, Phys.~Rev.~D {\bf 15} (1977) 2991.
\bibitem{Hofsaess78}T.~Hofs\"ass and G.~Schierholz, Phys.~Lett.~B {\bf 76}
(1978) 125.
\bibitem{Roth79}R.~Roth, Nucl.~Phys.~B {\bf 154} (1979) 21.
\bibitem{Roberts94}C.~D.~Roberts and A.~G.~Williams,
Prog.~Part.~Nucl.~Phys.~{\bf 33} (1994) 477, hep-ph/9403224.
\bibitem{Roberts00a}C.~D.~Roberts and S.~M.~Schmidt,
Prog.~Part.~Nucl.~Phys.~{\bf 45} (2000) S1, nucl-th/0005064.
\bibitem{Roberts00b}C.~D.~Roberts, nucl-th/0007054.
\bibitem{Alkofer01}R.~Alkofer and L.~von Smekal, Phys.~Rep.~{\bf 353} (2001)
281, hep-ph/0007355.
\bibitem{Maris02}P.~Maris, A.~Raya, C.~D.~Roberts, and S.~M.~Schmidt,
Eur.~Phys.~J.~A {\bf 18} (2003) 231, nucl-th/0208071.
\bibitem{Maris03}P.~Maris and C.~D.~Roberts, Int.~J.~Mod.~Phys.~E {\bf 12}
(2003) 297, nucl-th/0301049.
\bibitem{Roberts03}C.~D.~Roberts, Lect.~Notes Phys.~{\bf 647} (2004) 149,
nucl-th/0304050.
\bibitem{Finger82}J.~R.~Finger and J.~E.~Mandula, Nucl.~Phys.~B {\bf 199}
(1982) 168.
\bibitem{LeYaouanc84}A.~Le Yaouanc, L.~Oliver, O.~P\`ene, and J.-C.~Raynal,
Phys.~Rev.~D {\bf 29} (1984) 1233.
\bibitem{Adler84}S.~L.~Adler and A.~C.~Davis, Nucl.~Phys.~B {\bf 244} (1984)
469.
\bibitem{Kocic86}A.~Koci\'c, Phys.~Rev.~D {\bf 33} (1986) 1785.
\bibitem{Alkofer88}R.~Alkofer and P.~A.~Amundsen, Nucl.~Phys.~B {\bf 306}
(1988) 305.
\bibitem{Klevansky88}S.~P.~Klevansky and R.~H.~Lemmer, Phys.~Rev.~D {\bf 38}
(1988) 3559.
\bibitem{Alkofer89}R.~Alkofer, P.~A.~Amundsen, and K.~Langfeld, Z.~Phys.~C
{\bf 42} (1989) 199.
\bibitem{Babutsidze98}T.~Babutsidze, T.~Kopaleishvili, and A.~Rusetsky,
Phys.~Lett.~B {\bf 426} (1998) 139, hep-ph/9710278.
\bibitem{Babutsidze99}T.~Babutsidze, T.~Kopaleishvili, and A.~Rusetsky,
Phys.~Rev.~C {\bf 59} (1999) 976, hep-ph/9807485.
\bibitem{Kopaleishvili01}T.~Kopaleishvili, Phys.~Part.~Nucl.~{\bf 32} (2001)
560, hep-ph/0101271.
\bibitem{Babutsidze03}T.~Babutsidze, T.~Kopaleishvili, and D.~Kurashvili,
Georgian Electronic Scientific J.: Phys.~{\bf 1} (2004) 20, hep-ph/0308072.
\bibitem{Lucha05}Li Z.-F., W.~Lucha, and F.~F.~Sch\"oberl, in preparation.
\end{thebibliography}
\end{document}